\documentclass[a4paper]{article}

\usepackage{INTERSPEECH2021}
\usepackage{blindtext,lipsum,letltxmacro,xcolor,xparse}
\usepackage{comment}
\usepackage{hyperref,cite}
\usepackage{pifont}

\usepackage{amssymb}
\usepackage{pifont}

\makeatletter
               {\list{}{\leftmargin=10pt 
                        \labelwidth\z@ \itemindent-\leftmargin
                        }}%
               {\endlist}
\makeatother

\LetLtxMacro{\blindtextblindtext}{\blindtext}
\LetLtxMacro{\blindtextBlindtext}{\Blindtext}

\RenewDocumentCommand{\blindtext}{O{\value{blindtext}}}{%
  \begingroup\color{gray}\blindtextblindtext[#1]\endgroup
}
\RenewDocumentCommand{\Blindtext}{O{\value{blindtext}}O{\value{Blindtext}}}{%
  \begingroup\color{gray}\blindtextBlindtext[#1][#2]\endgroup
}

\def\y{{\mathbf y}}
\def\x{{\mathbf x}}

\def\xhat{{\hat{\mathbf x}}}
\def\i{{\mathbf i}}
\def\o{{\mathbf o}}
\def\a{{\mathbf a}}
\def\e{{\mathbf e}}
\def\eo{{\mathbf{e}^{\text{1-hot}}}}
\def\W{{\mathbf W}}
\def\ea{{\mathbf{e}^{\text{enrl}}}}

\title{Few-shot learning of new sound classes for target sound extraction}
\name{Marc Delcroix, Jorge Bennasar Vázquez, Tsubasa Ochiai, Keisuke Kinoshita, Shoko Araki}
\address{NTT Corporation, Japan }
\email{marc.delcroix@ieee.org}

\begin{document}
\setlength{\abovedisplayskip}{1pt}
\setlength{\belowdisplayskip}{1pt}
\setlength{\abovedisplayshortskip}{1pt}
\setlength{\belowdisplayshortskip}{1pt}
\maketitle
\begin{abstract}
Target sound extraction consists of extracting the sound of a target acoustic event (AE) class from a mixture of AE sounds. 
It can be realized using a neural network that extracts the target sound conditioned on a 1-hot vector that represents the desired AE class. With this approach, embedding vectors associated with the AE classes are directly optimized for the extraction of sound classes seen during training. However, it is not easy to extend this framework to new AE classes, i.e. unseen during training.
Recently, speech, music, or AE sound extraction based on enrollment audio of the desired sound offers the potential of extracting any target sound in a mixture given only a short audio signal of a similar sound. 
In this work, we propose combining 1-hot- and enrollment-based target sound extraction, allowing optimal performance for seen AE classes and simple extension to new classes. In experiments with synthesized sound mixtures generated with the Freesound Dataset (FSD) datasets, we demonstrate the benefit of the combined framework for both seen and new AE classes. Besides, we also propose adapting the embedding vectors obtained from a few enrollment audio samples (few-shot) to further improve performance on new classes.

\end{abstract}

\section{Introduction}
Human beings can focus on listening to a desired sound in a mixture, which enables us to follow our interlocutor despite others' conversations, to carefully listen to the violin part in a concert, or to pick up important acoustic events (AE) sounds such as a klaxon at a crossing. 
It has been a long-standing goal of researchers to reproduce human listening capabilities. 
Recently, neural network-based target sound extraction has received increased interest as a promising approach towards this goal, with methods developed to extract speech of a target speaker~\cite{zmolikova2017speaker,zmolikova2019speakerbeam,wang2018deep,wang2019voicefilter}, music instruments~\cite{slizovskaia2021conditioned,LeeCL19}  or AE sounds~\cite{Ochiai2020,gfeller2020oneshot,kong2020source}. In this paper, we focus on the AE sound extraction problem, which is particularly challenging given the large variety of sounds it covers (e.g. knock, telephone rings, cough, animal sounds, etc).

AE sound extraction consists of extracting the sound of a target AE class from a mixture of AE sounds. It uses an extraction neural network that estimates the target AE sound given the sound mixture and an embedding vector that represents the characteristics of the target sound. 
The embedding vector can be obtained using an embedding encoder that receives either (1) an \emph{enrollment audio sample} that is similar to the target AE sound~\cite{gfeller2020oneshot} or, (2) a \emph{1-hot vector} that represents the target AE class~\cite{kong2020source,Ochiai2020}. 
The extraction neural network and the embedding encoders are jointly trained.

The enrollment-based approach does not explicitly assume well-defined AE classes and thus can naturally handle target sounds from \emph{new classes} (i.e. unseen during training), given a short enrollment audio sample of a sound of that new class. However, the embedding vectors may not be optimal for certain AE classes if the AE class has relatively large intra-class variability e.g. animal sounds or phone rings, which may cause a mismatch between the enrollment and the target sound in the mixture.
In contrast, the 1-hot vector provides a way to optimize directly the embedding for the sound classes during training. It can thus achieve optimal performance on the seen AE classes. However, it cannot handle new AE classes.

In this paper, we propose a method to combine the advantages of both approaches by designing a \emph{mixed model} that can perform both 1-hot- and enrollment-based target sound extraction.
Since both approaches use embedding vectors, by mapping these embedding vectors to a common space, we can use either 1-hot- or enrollment-based approaches to extract a target sound.
This is realized by training with a multi-task scheme, i.e. training a shared extraction network simultaneously with embedding vectors obtained from 1-hot vectors and enrollment samples. 
With such a mixed model, we can directly optimize the embedding vectors for seen classes with the 1-hot encoder, while enabling the handling of new AE classes with the enrollment encoder. Besides, the regularization effect of the multi-task training scheme can also help to learn better models.

In experiments using mixtures generated from sound events taken from the Freesound Datasets (FSDs)~\cite{fonseca2018general,FonsecaPFFBFOPS17}, we confirm that the proposed mixed model can achieve superior performance on the extraction of seen AE classes compared to the baseline 1-hot- or enrollment-based approaches. Moreover, it can also generalize to new classes as the enrollment-based approach. We further investigate an adaptation/retraining strategy that enables to boost performance on new AE classes given a few samples from the new AE classes. The proposed method can thus learn 1-hot-based target sound extraction with a few labeled data of the new AE classes, which can be seen as few-shot learning.

We base our investigation using a network configuration similar to that of SpeakerBeam~\cite{delcroix2020improving}, which we proposed for enrollment-based target speech extraction. We thus refer to the target sound extraction framework as SoundBeam. In the remainder of the paper, we first describe the SoundBeam framework and the proposed mixed model in Section~\ref{sec:SoundBeam}. In Section~\ref{sec:new_classes}, we discuss how to handle new classes and introduce the adaptation framework. We introduce related works in  Section~\ref{sec:related_works}. In Section~\ref{sec:expe}, we present experimental results on seen and new AE classes. Finally, we conclude the paper in Section~\ref{sec:conclusion}.

\section{SoundBeam}

\label{sec:SoundBeam}
\begin{figure}[tb]
    \centering
    \includegraphics[width=0.4\textwidth]{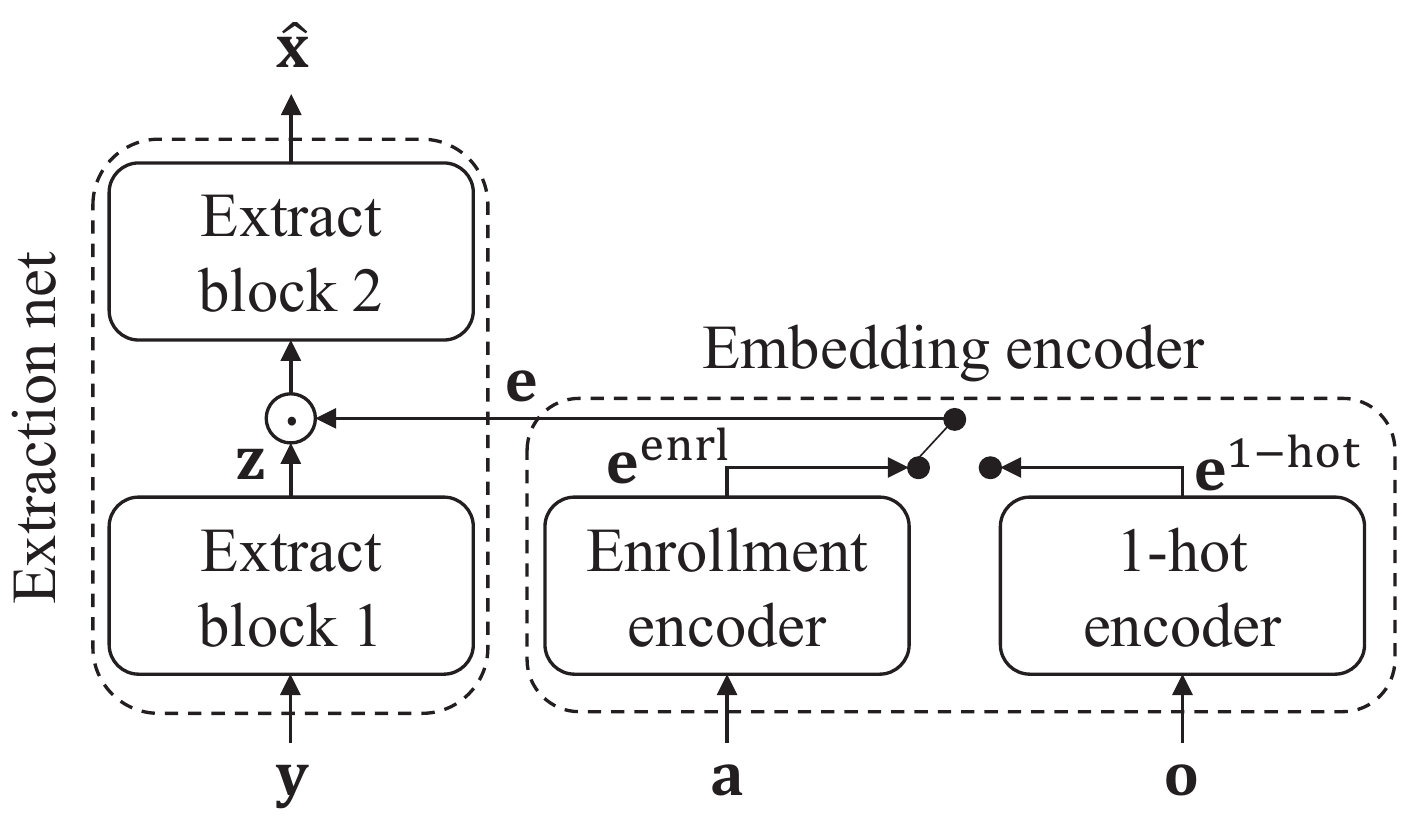}
    \vspace{-4mm}
    \caption{Diagram of SoundBeam showing the two  ways to obtain the embedding vectors, i.e. 1-hot- or enrollment-based.}
    \vspace{-6mm}
    \label{fig:method}
\end{figure}

We consider the problem of extracting sounds of a target AE class from a mixture of AE sounds given as,
\begin{align}
    \y = \x + \i,
\end{align}
where $\y \in \mathbb{R}^{T}$ and $\x \in \mathbb{R}^{T}$ are the observed single-channel mixture and the target sound signals, respectively. $\i \in \mathbb{R}^{T}$ is a signal that includes the interference signals from other AE classes and background noise, and $T$ is the signal duration.

Figure~\ref{fig:method} is a schematic diagram of SoundBeam. It consists of two networks (1) an \emph{extraction network} and (2) an \emph{embedding encoder} network.
The extraction network estimates the target sound signal, $\xhat \in \mathbb{R}^{T}$, as,
\begin{align}
 \xhat = f(\y, \e),   
\end{align}
where $f(\cdot)$ is a neural network, $\e \in \mathbb{R}^{D \times 1}$ is an embedding vector that represents the characteristics associated with the target AE class, and $D$ is the dimension of the embedding vector.

There are different ways to exploit the embedding vector within the extraction network~\cite{zmolikova2019speakerbeam,zeghidour2020wavesplit,gfeller2020oneshot}. We use here an element-wise multiplication ($\odot$ operator in Fig.~\ref{fig:method}) between the internal representation of the mixture signal after the first extraction block, $\mathbf{z}$, and the embedding vector, $\e$.
Note that the extraction network could directly predict the target sound signal or a mask that is applied to the mixture to extract the target sound signal as in~\cite{delcroix2020improving}. We use the mask-based approach in this study.

The embedding encoder computes the embedding vector $\e$. 
There are two ways of estimating the embedding vector: using a 1-hot vector~\cite{kong2020source,Ochiai2020} or an enrollment audio sample~\cite{gfeller2020oneshot}. We describe these two approaches in the following subsections.
 
\subsection{1-hot-based SoundBeam}
\label{ssec:1-hot SoundBeam}
Let $\o=[o_1, \ldots, o_N]^T$ be a 1-hot vector, which characterizes the AE class of the target sound,
i.e. a $N\times 1 $ vector with $i$-th element given as $o_i = \delta_{i,n}$,
where $\delta_{i,n}$ is the Kronecker delta, 
$n$ is the index of the target AE class,
and $N$ is the number of AE classes.
The 1-hot encoder consists of an embedding layer.
The embedding vector obtained from the 1-hot vector, $\eo \in \mathbb{R}^{D \times 1}$, is given by,
\begin{equation}
    \eo = \W  \o,
\end{equation}
where  $\W \in \mathbb{R}^{D \times N}$ is an embedding matrix, 
whose columns contain the embedding vectors for each AE class.

The embedding layer can be trained jointly with the extraction network~\cite{kong2020source,Ochiai2020}, which enables to directly optimize the embedding vectors $\eo$ for each AE sound class. As a side-effect, the AE sound classes that the method can handle are fixed by the embedding matrix $\W$ and are limited to the AE sound classes seen during training. After training, it is thus not straightforward to handle new AE classes. 

\subsection{Enrollment-based SoundBeam}
\label{enrl SoundBeam}
Another approach to obtain the embedding vector is to use an enrollment audio sample, $\a \in \mathbb{R}^{T^a}$, of a sound similar to the target sound.
Note that in general the duration of the enrollment signal $T^a$ may differ from the duration of the mixture $T$.
The embedding derived from the enrollment sample, $\ea \in \mathbb{R}^{D \times 1}$, can be obtained as,
\begin{align}
    \ea = g(\a),
\end{align}
where $g(\cdot)$ represents the enrollment encoder, which is a neural network that maps the enrollment signal to a $D\times 1$ vector using e.g. average pooling across time-axis. 

The embedding-based approach does not directly optimize the embedding vector for the AE classes, which can result in lower performance for seen AE classes.
However, the method can naturally handle new AE classes, when we provide an enrollment sample with similar sound characteristics as the target, and if the system has been trained with a sufficient variety of AE sounds~\cite{gfeller2020oneshot}.

\subsection{Training a shared embedding space: SoundBeam-mixed}
The above formulation of target sound extraction emphasizes the similarities of the 1-hot- and enrollment-based approaches. Both approaches represent AE sounds in an embedding space. However, when trained separately, the 1-hot- and enrollment-based approaches will lead to different representations of the AE sounds. 
We propose to map the embedding vectors obtained by both approaches to a common embedding space so that we can enjoy the advantages of both frameworks, i.e. optimal performance on seen AE classes and generalization to new AE classes. 

We realize this by creating a model (called \emph{SoundBeam-mixed}) that includes both the enrollment and the 1-hot encoders and a shared extraction network.
SoundBeam-mixed is trained with multi-task (or multi-branch) training, i.e. we alternate between the 1-hot and the enrollment encoders during training. 
This will ensure that both the 1-hot and enrollment encoders provide a similar representation of AE sounds. 
Besides combining the strength of both approaches, we also expect improved extraction performance with SoundBeam-mixed thanks to the regularization effect of multi-task learning.

The training loss of \emph{SoundBeam-mixed} consists of,
\begin{align}
    \mathcal{L} =& \mathcal{L}^{ext}\left(\xhat=f(\y, \ea), \x \right) +\mathcal{L}^{ext}\left(\xhat=f(\y, \eo), \x \right) \nonumber \\
    & + \alpha \mathcal{L}^{emb}(\ea,\eo),
    \label{eq:mt_loss}
\end{align}
where $\mathcal{L}^{ext}(\cdot)$, $\mathcal{L}^{emb}(\cdot)$ and  $\alpha$ are an extraction loss, an embedding loss (EL) and a weighting parameter, respectively. Note that the other versions of SoundBeam are trained using only an extraction loss.
We use the negative scale-dependent signal-to-noise ratio (SNR)~\cite{roux2019sdr} as the extraction loss,
\begin{align}
\mathcal{L}^{ext} (\xhat, \x )= - 10 \log_{10} \biggl( \frac{\| \mathbf{x} \|^{2}}{\| \mathbf{x} - \hat{\mathbf{x}} \|^{2}} \biggr)
\end{align}
Various EL could be used. Here we simply compute the distance between the embedding vectors of both encoders as,
\begin{align}
\mathcal{L}^{emb} = d(\ea, \eo),
\end{align}
where $d(\cdot)$ is the cosine distance between the embedding vectors.

Note that by simply sharing the extraction network parameters, 
we already constrain both encoders to map their inputs into a common embedding space. However, the additional embedding loss (EL) may further help the embedding spaces becoming more similar.
The multi-task loss proposed in Eq.(\ref{eq:mt_loss}) may bring a mutual regularization effect, i.e. the 1-hot encoder may help the enrollment encoder learn more general sound embedding vectors, while the enrollment encoder may help the 1-hot encoder capture sound characteristics of the AE classes.

\section{Handling new classes}
\label{sec:new_classes}
For practical applications, it is important to be able to flexibly extend a sound extraction system to new AE classes after deployment using only a few audio samples from the new classes.
Let us assume that we have $K$ enrollment audio samples $\{\a_1,\ldots, \a_K\}$ from a new AE class. 
Thanks to the enrollment encoder, enrollment-based SoundBeam or mixed SoundBeam can naturally handle new classes.

For example, to extract AE sounds from a new class, we can average the $K$ embedding vectors computed with the enrollment encoder as,
\begin{align}
    \e^{new} = \frac{1}{K} \sum_{k=1}^K g(\a_k).
    \label{eq:avg_emb}
\end{align}
We can then use $\e^{new}$ as the embedding vector for the new AE class.

To further improve the quality of the embedding vectors, we propose to directly adapt the new embedding matrix vectors.
To do that, we first register the new classes by adding $\e^{new}$ to the embedding matrix as,
\begin{align}
    \W' = [\W, \e^{new}],
\end{align}
where $\W' \in \mathbb{R}^{D \times (N+1)}$ is a new embedding matrix.
We then retrain the embedding vector of the new class by using adaptation data.
Here, we create an adaptation dataset by mixing sounds from the training data and the $K$ enrollment audio samples. Note that, if we fix all network parameters but the new embedding, $\e^{new}$, we can adapt the embedding vector for the new AE class while keeping the previous embedding vectors unchanged, which will guarantee a constant level of performance for the AE classes seen during training.

\section{Related works}
\label{sec:related_works}
Target sound extraction can be realized by combining sound separation~\cite{kavalerov2019universal,tzinis2020improving} and event identification~\cite{komatsu2016acoustic,jeon2017nonnegative,Huang2020,dcase2020}. 
However, separation approaches require knowing or estimating the number of sources.
Moreover, a single model as SoundBeam can be jointly optimized, which can lead to superior performance~\cite{Ochiai2020}.

The 1-hot-based SoundBeam described in section~\ref{ssec:1-hot SoundBeam} corresponds to the universal sound extractor~\cite{Ochiai2020}. Concurrently~\cite{kong2020source} proposed a similar 1-hot-based sound extraction system that combines a sound event detection with an extraction network, to allow training on weakly labeled data consisting of audio clips containing several AEs without labels for their time occurrence.

The enrollment-based SoundBeam follows a similar implementation as SpeakerBeam~\cite{delcroix2020improving}, but is applied to AE sounds instead of speech. It is also conceptually similar to~\cite{gfeller2020oneshot,LeeCL19} but with different network architectures and different training schemes. Both works assumed AE class labels were not available during training and assured similar sound characteristics for the target enrollment by using either a different portion of the same sound sample~\cite{gfeller2020oneshot}, or the same sample for enrollment and target sound during training~\cite{LeeCL19}. 
Moreover,~\cite{LeeCL19} proposed converting the model to a 1-hot-based model using averaging of the embedding of the same class in the training set when labels were available, similar to Eq. (\ref{eq:avg_emb}). However, this approach does not directly optimize the embedding vectors as our proposed SoundBeam-mixed does. Besides, it was tested only for music sound extraction, with a small number of seen classes.

Compared to~\cite{gfeller2020oneshot,LeeCL19,kong2020source}, the proposed SoundBeam-mixed requires labeled data to train the 1-hot encoder. However, we could exploit both labeled and unlabelled data by borrowing the idea of~\cite{gfeller2020oneshot} to augment the data to train the enrollment encoder. Besides, we could also use a similar approach to~\cite{kong2020source,pishdadian2020learning} to generate labels to train the 1-hot encoder with weakly-labeled training data. These will be part of our future works.

\section{Experiments}
\label{sec:expe}
\subsection{Dataset}
To evaluate the effectiveness of the proposed method, we created a dataset of simulated sound event mixtures based on the FSD-Kaggle 2018~\cite{fonseca2018general} and the FSD50K corpora~\cite{FonsecaPFFBFOPS17}.

We created mixtures by mixing 3 AE sounds randomly selected from different AE classes. We included stationary background noise to the mixtures at a signal-to-noise ratio (SNR) between 15 and 25 dB, using noise samples from the  REVERB challenge corpus (REVERB)~\cite{kinoshita2016summary}.
All mixtures are six-second long and were generated by randomly extracting three audio clips of 2 to 5 seconds from the FSD corpus and pasting (adding) them to random time-positions on top of the six-second background noise.
We created sound event mixtures by utilizing Scaper~\cite{salamon2017scaper}.
In this experiment, we downsampled the sounds to 8 kHz to reduce the computational and memory costs. All experiments were performed with mixtures of 3 AE classes, but We confirmed in~\cite{Ochiai2020} that SoundBeam could handle mixtures with more AE classes. 

The training and development sets consist of 50,000 and 10,000 mixtures, respectively.
We used sound samples randomly selected from the 41 AE classes from the training set of the FSD-Kaggle dataset. These include AEs such as human sounds, object sounds, musical instruments, etc~\cite{fonseca2018general}.
We generated two test sets using the FSD-Kaggle and a subset of the FSD50K data that consists of the AE sound samples from a single AE class provided in the FUSS dataset~\cite{wisdom2020FUSS}. We used FSD50K to generate data with new AE classes unseen in the FSD-Kaggle training set.
The first test set (\emph{mixtures of seen AE classes}) consists of 10000 mixtures of sounds from the AE classes seen during training generated from FSD-Kaggle and is the same as that used in~\cite{Ochiai2020}. 
The second test set (\emph{mixtures with new AE classes}) consists of 3000 mixtures of sounds from two seen classes and one new AE class out of 10 new AE classes (including male speech, electric guitar, camera, etc). It is generated by using both FSD-Kaggle and FSD50K.

For the enrollment-based experiments on seen classes, we randomly selected an enrollment audio sample from the AE class of the target AE sound that differs from the target sound.
For the experiment on new classes, we randomly selected $K$ enrollment utterances to compute the average embedding vectors as shown in Eq. (\ref{eq:avg_emb}). We performed experiments with $K{=}1,5,10$.
We also created an adaptation set, which consisted of 1000 mixtures generated by mixing sounds from 2 AE classes from the training set and 1 sound from the 10 new AE classes, randomly selected from the $K$ enrollment utterances. 
Note that to simplify the experiments, we performed adaptation simultaneously for 10 new AE classes. However, since we only update the embedding matrix, it is equivalent to performing adaptation for each new AE class at a time.


\subsection{Experimental settings}

For all the experiments, we adopted a Conv-TasNet-based network architecture, which consists of stacked dilated convolution blocks.
We used the Asteroid toolkit for all experiments~\cite{Pariente2020Asteroid}.
By following the notations of~\cite{luo2019conv}, we set the hyper-parameters as follows: $N{=}256$, $L{=}20$, $B{=}256$, $H{=}512$, $P{=}3$, $X{=}8$, and $R{=}4$.
We also set the dimension of the embedding vectors to $D{=}256$.
The embedding vector is multiplied with the output of the first stacked convolution block (Figure~\ref{fig:method}). 
The enrollment encoder block consists of one stacked dilated convolution block ($R{=}1$) with the other hyper-parameters as for the extraction network.
We used the Adam algorithm~\cite{kingma2015adam} for optimization with an initial learning rate of $10^{-4}$ and used gradient clipping~\cite{pascanu2013difficulty}.
For the experiments with EL, we set $\alpha=3$.
All models were trained for up to 200~epochs, and we used the models achieving the lowest cross-validation loss value in all experiments. 
During adaptation, we fixed all network parameters except for the new embedding vectors. We retrained the new embedding vectors for 10 epochs with a learning rate of~$10^{-3}$.

We evaluate the results in terms of scale-invariant signal-to-distortion ratio (SDR) computed with the BSSEval toolkit~\cite{vincent2006performance}.
The results were obtained by averaging the SDR for each AE sound in each mixture of the test sets.

\subsection{Results on mixtures of seen AE classes}
\begin{table}[t]
  \caption{SDR improvement [dB] for experiment with mixtures of seen AE classes. The SDR of the mixture was -3.6 dB.}
\vspace{-3mm}
  \label{tab:result_seen}
  \centering
  \begin{tabular}{ l c c c c }
\toprule
\bf{Embedding}         & \multicolumn{4}{c}{\bf{Model}} \\
\bf{at test time} & 1-hot &	Enrl & Mixed & Mixed+EL \\
\midrule
1-hot & 11.4 & - & 12.6 & \bf{12.9} \\
Enrl & - & 10.4 & \bf{10.5} & 10.1 \\
\bottomrule
  \end{tabular}
\vspace{-1mm}
\end{table}
Table~\ref{tab:result_seen} shows the SDR improvement for the test set consisting of seen AE classes. The table compares results with 1-hot (``1-hot'') and enrollment (``Enrl'')-based SoundBeam and the proposed SoundBeam-mixed, with and without EL, when using 1-hot- or enrollment-based embedding vectors at test time. 1-hot- and enrollment-based models serve as baselines for target sound extraction
~\cite{gfeller2020oneshot,Ochiai2020}.

The results confirm that the 1-hot based approach, which performs direct optimization of the embedding vectors, outperforms the enrollment-based approach. 
The proposed SoundBeam-mixed models further improve SDR by more than 1 dB.
The model trained with EL achieves the best performance when using the 1-hot encoder.
These results demonstrate the positive effect of the proposed multi-task training scheme on the seen classes.

\subsection{Results on mixtures with new AE classes}

\begin{table}[t]
  \caption{SDR improvement [dB] for experiment on mixtures with new AE classes. The SDR for the mixture signals was $-3.4$ dB for the seen AEs and $-4.0$ dB for the new AEs.
  We used 1-hot-based encoder for extracting seen AEs except for the enrollment-based model, and Eq. (\ref{eq:avg_emb}) to compute the embedding vectors for the new AE classes.
  }
\vspace{-3mm}
  \label{tab:result_new}
  \centering
  \begin{tabular}{l   c ccc }
\toprule
\bf{Model} &     \bf{Seen AEs} & \multicolumn{3}{c}{\bf{New AEs}} \\
        &            & $K{=}1$ & $K{=}5$ & $K{=}10$ \\
    \midrule
1-hot  & 	10.5&	-&	-&	- \\
\midrule
Enrl & 10.4	&\bf{4.9}&	7.0	&7.0 \\
\midrule
Mixed  &     	11.6&	4.4	&7.2 & 7.5	\\
\midrule
Mixed + EL & 	\bf{11.8} &	3.2	&7.1&	7.4\\
+adapt (rnd init)  & - & 0.2 &3.4	& 5.1 \\
+adapt (avg init)	& -	&3.8 &	\bf{7.8}	& \bf{8.2} \\
\bottomrule
  \end{tabular}
\vspace{-6mm}
\end{table}
Table~\ref{tab:result_new} shows the SDR improvement for mixtures with new AE classes for the 1-hot, enrollment, and mixed models without and with adaptation. 
The left part of the table shows the extraction performance of the two seen AE classes in the mixtures. For the extraction of seen AE classes, we used 1-hot embedding vectors for all models except for the enrollment-based model. These results demonstrate that all models can extract sounds of the seen classes even if the mixtures include other sounds unseen during training.

The right-side of Table~\ref{tab:result_new} shows the SDR improvement for the extraction of sounds from new AE classes for different numbers of enrollment samples. We used Eq.~(\ref{eq:avg_emb}) to compute the embedding vectors for the new classes.
The results demonstrate that the proposed mixed models also improve extraction performance for the new classes when $K>1$. 
With a single enrollment sample ($K{=}1$), the sample may not well represent the new AE class causing a mismatch between the enrollment and the target sound. Consequently, the average extraction performance is relatively low.
However, with more enrollment samples ($K{=}5, 10$) extraction performance greatly improves especially with SoundBeam-mixed, which achieves an SDR improvement of up to 7.5 dB. 

Finally, the last rows of table~\ref{tab:result_new} show the effect of adaptation (``+adapt'') by retraining the embedding vectors associated with the new AE classes (see section~\ref{sec:new_classes}). We considered two configurations for adaptation, with embedding randomly initialized (``rand init'') or initialized with the averaged embedding vectors as shown in Eq.~(\ref{eq:avg_emb}) (``avg init'').
Adaptation from randomly initialized embedding vectors achieves an SDR improvement of up to about 5.1~dB, which is less than the 7.4~dB improvement obtained with the mixed model with averaged embedding vectors without adaptation.
Adapting the averaged embedding vectors improves performance compared by up to 0.8~dB.

\begin{figure}[tb]
    \centering
    \includegraphics[width=0.45\textwidth]{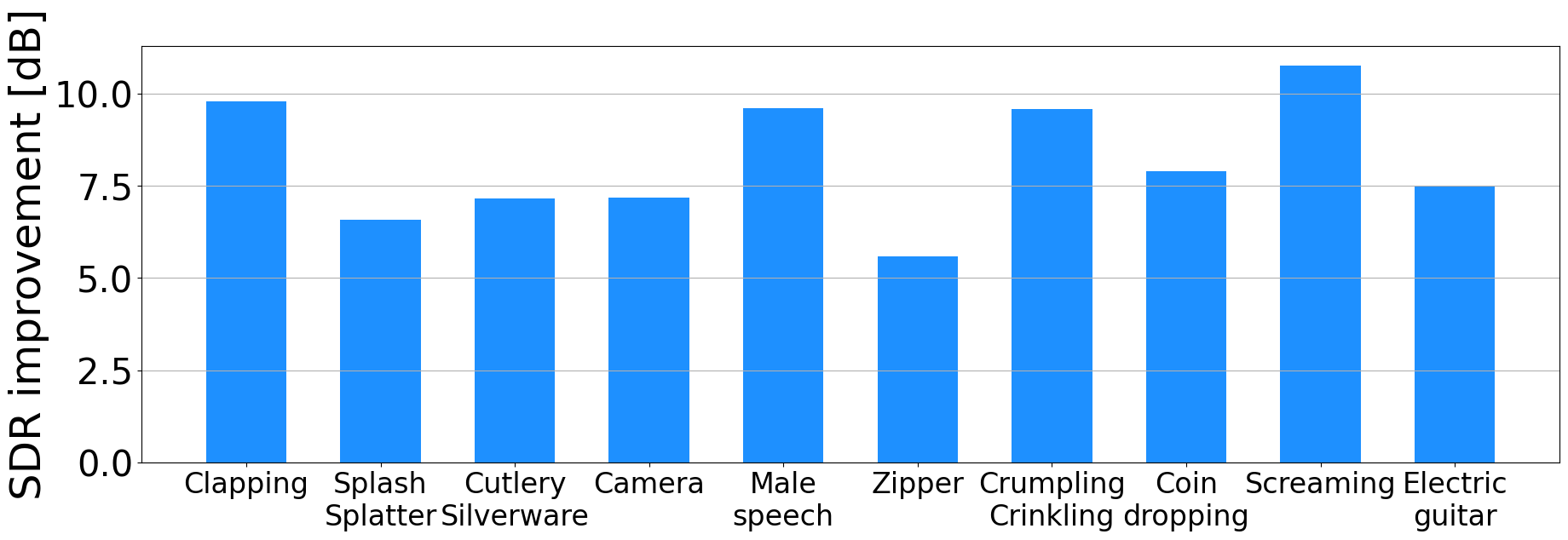}
    \vspace{-3mm}
    \caption{SDR improvement for SoundBeam-mixed with EL and adaptation for the 10 new AE classes with $K{=}10$.}
    \vspace{-6mm}
    \label{fig:sdr_imp}
\end{figure}
Figure~\ref{fig:sdr_imp} plots the SDR improvement for the new AE classes with the proposed SoundBeam-mixed with adaptation. The proposed scheme could achieve an SDR improvement of more than 5 dB on all 10 new classes. This result demonstrates that the proposed scheme 
can realize efficiently few-shot learning of target sound extraction for new AE classes.

\section{Conclusion}
\label{sec:conclusion}
We have proposed a novel target sound extraction system that combines both 1-hot- and enrollment-based approaches.
We showed that by designing a sound extraction model with both 1-hot and enrollment encoders and a shared extraction network, we can enforce both approaches to share a common embedding space. This mixed model improves significantly extraction performance thanks to multi-task training. Moreover, it enables 1-hot based target sound extraction on new classes. Combined with retraining-based adaptation, we could achieve high extraction performance on new AE classes with only a few samples. 

In future works, we will consider training the system on a larger dataset, which may include audio clips with multiple AE classes, or unlabeled data~\cite{gfeller2020oneshot,kong2020source}. Moreover, we will also investigate other EL~\cite{zeghidour2020wavesplit} to provide more discriminative embedding vectors and extend the model to perform online processing~\cite{tzinis2021compute}.

\bibliographystyle{IEEEtran}
\bibliography{mybib}

\end{document}